\documentstyle[12pt]{article}

\def\lsim{\;\raise0.3ex\hbox{$<$\kern-0.75em\raise-1.1ex\hbox{$\sim$}}\;}
\def\gsim{\;\raise0.3ex\hbox{$>$\kern-0.75em\raise-1.1ex\hbox{$\sim$}}\;}

\makeatletter

\def\section{\@startsection {section}{1}{\z@}{-3.5ex plus -1ex minus
    -.2ex}{5ex plus .2ex}{\Large\centering\bf}}
\def\subsection{\@startsection{subsection}{2}{\z@}{-3.25ex plus -1ex minus
   -.2ex}{3ex plus .2ex}{\large\centering\bf}}
\def\subsubsection{\@startsection{subsubsection}{3}{\z@}{-3.25ex plus
 -1ex minus -.2ex}{1.5ex plus .2ex}{\normalsize\centering\bf}}
\def\paragraph{\@startsection
     {paragraph}{4}{\z@}{3.25ex plus 1ex minus .2ex}{-1em}{\normalsize\bf}}
\def\subparagraph{\@startsection
     {subparagraph}{4}{\parindent}{3.25ex plus 1ex minus
     .2ex}{-1em}{\normalsize\em}}

\def\maketitle{\par
 \begingroup
   \def\thefootnote{\fnsymbol{footnote}}
   \def\@makefnmark{\hbox
       to 0pt{$^{\@thefnmark}$\hss}}
   \if@twocolumn
     \twocolumn[\@maketitle]
     \else \newpage
     \global\@topnum\z@        % Prevents figures from going at top of page.
     \@maketitle \fi\thispagestyle{plain}\@thanks
 \endgroup
 \setcounter{footnote}{0}
 \let\maketitle\relax
 \let\@maketitle\relax
 \gdef\@thanks{}\gdef\@author{}\gdef\@title{}\let\thanks\relax}

\def\@maketitle{\newpage
 \null
 \vskip 2em                 % Vertical space above title.
 \begin{center}
  {\LARGE \@title \par}     % Title set in \LARGE size.       %%%
  \vskip 1.5em                % Vertical space after title.
  {\large                       % each author set in \large, in a   %%%
   \lineskip .5em           % tabular environment
   \begin{tabular}[t]{c}\@author
   \end{tabular}\par}
  \vskip 1em              % Vertical space after author.
  {\large \@date}           % Date set in \large size.    %%%
\end{center}
 \par
 \vskip 1.5em}                % Vertical space after date.

\def\abstract{\if@twocolumn
\section*{Abstract}
\else \small
\begin{center}
{\bf Abstract\vspace{-.5em}\vspace{0pt}}
\end{center}
\quotation
\fi}

\def\endabstract{\if@twocolumn\else\endquotation\fi}

 \newlength{\captionwidthh}

 \long\def\@caption#1[#2]#3{\addcontentsline{\csname
  ext@#1\endcsname}{#1}{\protect\numberline{\csname
  the#1\endcsname}{\ignorespaces #2}}\par
    \begingroup
      \setlength{\captionwidthh}{\textwidth}
      \addtolength{\captionwidthh}{-2cm}
      \begin{center}
      \parbox{\captionwidthh}{
        \small
        \@makecaption{\csname fnum@#1\endcsname}{\ignorespaces #3}\par
      }
      \end{center}

    \endgroup }

\def\@cite#1{#1}                % Produces the output of the \cite command.

\def\thebibliography#1{\section*{References\@mkboth
  {REFERENCES}{REFERENCES}}\list
  {[\arabic{enumi}]}{\settowidth\labelwidth{#1}\leftmargin\labelwidth
    \advance\leftmargin\labelsep
    \addtolength\itemindent{-\labelwidth}
    \usecounter{enumi}}
    \def\newblock{\hskip .11em plus .33em minus .07em}
    \sloppy\clubpenalty4000\widowpenalty4000
    \sfcode`\.=1000\relax}

\def\@lbibitem[#1]#2{\item[]\if@filesw
      { \def\protect##1{\string ##1\space}\immediate
        \write\@auxout{\string\bibcite{#2}{#1}}}\fi\ignorespaces}
\makeatother

\newcommand{\pspict}[2]{
  \vspace*{#1}
  \special{dvitops: import #2 \the\textwidth #1}
}

\newcommand{\zerotxt}[1]{
        \mbox{$\lefteqn{\mbox{#1}}$}
}

\newlength{\numpit}

\newlength{\indeksinpituus}

\newlength{\ypit}

\newcommand{\be}{\begin{equation}}
\newcommand{\ee}{\end{equation}}
\newcommand{\ba}{\begin{eqnarray}}
\newcommand{\ea}{\end{eqnarray}}

\newlength{\pohjaviiva}
\newcommand{\vsas}{\vspace*{0.0cm}}

\newcommand{\confps}[1]{
\vspace{7.5cm}
\special{dvitops: import #1 \the\textwidth 10cm}
\vspace{-3.5cm}
}

\begin{document}

%% Coverpages
%\input kansi.tex

\pagestyle{empty}

\setcounter{page}{-3}

\vspace*{-20mm}

\begin{center}
{\bf RESEARCH INSTITUTE FOR HIGH ENERGY PHYSICS}
\end{center}
\vspace*{1mm}
\begin{center}
\Large 
\bf REPORT SERIES
\end{center}
\vspace*{1.0mm}
\begin{center}

\noindent {\bf HU - SEFT \hspace{1.5 mm} R  \hspace{1.5 mm} 1996 - 07}

\vspace*{23mm}

{\huge\bf TESTS~OF~THE}

\vspace*{2mm}
{\huge\bf STANDARD~MODEL}

\vspace*{2mm}
{\huge\bf AND~ITS~EXTENSIONS}

\vspace*{2mm}
{\huge\bf IN~HIGH-ENERGY}

\vspace*{2mm}
{\huge\bf $\bf e^+e^-$~COLLISIONS}

\vspace*{35mm}

\vspace{6mm}

\noindent{\Large\bf RAIMO~~VUOPIONPER\"A}

\vspace*{6mm}

{\large\bf Research Institute for High Energy Physics}

\vspace*{-2mm} {\large\bf University of Helsinki}

\vspace*{-2mm} {\large\bf Helsinki, Finland}

\vspace*{13.82mm}

\vspace*{5mm}
\input epsf
\vspace{-0.12cm}
\begin{center}
\mbox{\epsfxsize=4.05cm\epsfysize=1.46cm\epsffile{./logo.seft}}
\end{center}

ISSN 0788-3587

\vspace{0.4 cm}
UNIVERSITY OF HELSINKI \\
\vspace*{-2mm}
RESEARCH INSTITUTE FOR HIGH ENERGY PHYSICS \\
\vspace*{-2mm}
P.O.Box 9 $\bullet$ FIN-00014 UNIVERSITY OF HELSINKI $\bullet$ FINLAND

\end{center}

\normalsize\rm

\newpage

\vspace*{20cm}

\begin{center}

\vspace*{-3.5mm}
{\footnotesize\bf
 ISSN 0788-3587}

\vspace*{-3.5mm}
{\footnotesize\bf
Helsinki University Press 1996}

\vspace*{-3.5mm}
{\footnotesize\bf
Pikapaino}
\end{center}

\newpage

\vspace*{-20mm}

\begin{center}
{\bf RESEARCH INSTITUTE FOR HIGH ENERGY PHYSICS}
\end{center}
\vspace*{1mm}
\begin{center}
\Large 
\bf REPORT SERIES
\end{center}
\vspace*{1.0mm}
\begin{center}

\noindent {\bf HU - SEFT \hspace{1.5 mm} R \hspace{1.5 mm} 1996 - 07}

\vspace*{23mm}

{\huge\bf TESTS~OF~THE}

\vspace*{2mm}
{\huge\bf STANDARD~MODEL}

\vspace*{2mm}
{\huge\bf AND~ITS~EXTENSIONS}

\vspace*{2mm}
{\huge\bf IN~HIGH-ENERGY}

\vspace*{2mm}
{\huge\bf $\bf e^+e^-$~COLLISIONS}

\vspace*{35mm}

\vspace{6mm}

\noindent{\Large\bf RAIMO~~VUOPIONPER\"A}

\vspace*{6mm}

{\large\bf Research Institute for High Energy Physics}

\vspace*{-2mm} {\large\bf University of Helsinki}

\vspace*{-2mm} {\large\bf Helsinki, Finland}

\vspace*{12mm}

{\small\it ACADEMIC DISSERTATION}
\smallskip

\vspace*{-2mm}
{\small\it
To be presented, with the permission of the Faculty of Science}

\vspace*{-2mm}
{\small\it  of the University of Helsinki, for public criticism in 
Auditorium XIV}

\vspace*{-2mm}
{\small\it on April 13th, 1996, at 10 o'clock.}

\vspace*{5mm}
\input epsf
\vspace{-0.12cm}
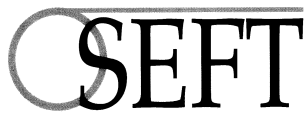
\begin{center}
\mbox{\epsfxsize=4.05cm\epsfysize=1.46cm\epsffile{./logo.seft}}
\end{center}

{\large\noindent Helsinki 1996}

\end{center}

\normalsize\rm

\newpage

\vspace*{20cm}

\begin{center}

\vspace*{-3.5mm}
{\footnotesize\bf
ISBN 951-45-7356-0}

\vspace*{-3.5mm}
{\footnotesize\bf
ISSN 0788-3587}

\vspace*{-3.5mm}
{\footnotesize\bf
Helsinki University Press 1996}

\end{center}

%% End Coverpages

\newpage

\pagestyle{plain}

\setcounter{page}{1}
\pagenumbering{arabic}

\setlength{\parskip}{0mm}

\vsas
\section*{Preface}
        \addtocontents{toc}{\protect\addvspace{3mm}}
        \addcontentsline{toc}{section}{Preface}
        \addtocontents{toc}{\protect\addvspace{3mm}}

This Thesis is based on research carried out
at the Department of High Energy Physics of University of Helsinki,
at the Research Institute for High Energy Physics (SEFT),
at Conseil Europeen pour la Recherche Nucleaire (CERN),
at the High Energy Physics  Laboratory of the Department of Physics of the  
University of Helsinki, and at the  Deutsches Elektronen-Synchrotron (DESY).
The work has been supported by the Lapin
rahasto of the Finnish Cultural Foundation and 
the Vilho, Yrj\"o and Kalle V\"ais\"al\"a Foundation. My stay at DESY was 
funded by the Province of Finnish Lapland, the City of Rovaniemi and 
the municipality of Rovaniemen maalaiskunta. 
I wish to express my gratitude to
these organizations, institutes and foundations.

\bigskip

I want to thank my supervisor and collaborator Docent Jukka 
Maalampi. His
guidance and teaching has been really inspiring
and he has given me the freedom to 
work with my own pace and on my own areas of interest.
I want to thank my physics and mathematics teacher in gymnasium,
Tapio Nygren, whose
excellent and demanding teaching aroused my interest to the wonderful
world of physics and particle physics.

\medskip

I am grateful to Docent Risto Orava, the director of SEFT, who 
has created an excellent research facility and given me the
opportunity to work at the Institute. He has also given me the
freedom which I needed and he has never restricted my research projects.
I also wish to thank Professor Masud Chaichian for his excellent teaching 
and many wonderful and useful conversations.

\medskip

Warm thanks are due to Dr. Reino Ker\"anen, Mr. Jari Pennanen and Dr.
Martti Raidal for
their co-authorship and discussions.  Also I would like to thank Docent Jukka 
Maalampi, Professors Kari Enqvist and Masud Chaichian and 
Dr. Kati Huitu for valuable 
comments and on careful reading of the manuscript. 

\medskip

I thank the colleagues and the personnel at
the physics institutes 
for making these first nine years of my academic life
an enjoyable period. Especially warm thanks are due to Drs. Ricardo Gonzalez
Felippe and  Martti Raidal for many memorable moments.

\medskip

Warm and loving thanks are due to my parents Petter and Estrid
and to my  brother Aarne and to my sister Linnea for their support and warmth.
I thank also all my friends for their friendship and understanding.

\bigskip

This Thesis is dedicated 
to freedom and 
to all 
open minded
human beings 
in our beautiful world.

\vspace{0.5cm}

\noindent\hspace*{113mm}\begin{minipage}[t]{42mm}Helsinki, March 5, 1996
\vspace*{0.2cm}\\
{Raimo Vuopionper\"a}\end{minipage}

\newpage

\setlength{\parskip}{2mm}

%% Abstract

\setlength{\pohjaviiva}{\baselineskip}
\setlength{\baselineskip}{6mm}
\noindent
{\bf Tests of the Standard Model and its extensions 
in high-energy $e^+e^-$ collisions}\\
\underline{Raimo} Helmeri Vuopionper\"a \\
University of Helsinki, 1996

\vspace{0.5cm}
\vspace*{-0.6cm}
\section*{Abstract}
        \addcontentsline{toc}{section}{Abstract}
        \addtocontents{toc}{\protect\addvspace{3mm}}

\vspace*{-0.5cm}

This thesis concerns the testing of the SM and its extensions
in $e^+e^-$ collisions, the main emphasis being on neutrino physics. 

The future $e^+e^-$ colliders will provide an excellent environment for
precision tests of the Standard Model (SM) of particle interactions, as well
as for a search of possible new phenomena going beyond the SM.
The LEP upgrade will make it possible to test the self-interactions of the
gauge bosons and other not so well known features of the SM, and at the planned
TeV-range linear colliders one would be able to explore various extended
schemes. 

Besides the recently discovered top quark and the so far undetected Higgs
particle, the tau neutrino ($\nu_\tau$) belongs to the most poorly known
constituents of the model.
A new concept for detecting the tau neutrino induced reactions in matter is
presented. It is based on a very asymmetric $e^+e^-$ collider combined with a
large coarsely instrumented detector. Applying the same beam parameters as
planned for the $500$ GeV linear colliders, it is found that the tau neutrino
would be detectable, though marginally, and its signature would be very clean.

The single top quark production at the LEP200 is also studied. While not
important at LEP200, this process will play an important role as a $t$-quark
source in the future $e^+e^-$ linear colliders.

The pair production of heavy neutrinos and the single heavy
neutrino production at $e^+e^-$ colliders are also studied, and methods for
distinguishing
heavy Dirac and Majorana neutrinos are presented. At the $500$ GeV  $e^+e^-$
linear collider the neutrinos with mass $m_N \lsim 150$ GeV would be clearly
detectable. It is shown that Dirac and Majorana neutrinos can be distinguished
by using the angular distributions of the production processes and the
production threshold behaviour of the heavy neutrino pair production.

The experimental implications of the left-right symmetric
$SU(2)_L\otimes SU(2)_R\otimes U(1)_{B-L}$ model
are also investigated. The single charged triplet Higgses $\Delta^\pm$
predicted by the left-right symmetric 
model are found to be observable at the upgraded $e^+e^-$ collider with 
$\sqrt{s} = 1 - 2$ TeV and ${\cal L}_{\rm year} \approx {\cal O}(100)$
${\rm fb}^{-1}$.

\clearpage
\setlength{\baselineskip}{\pohjaviiva}

\vsas
\tableofcontents
\newpage

\setlength{\baselineskip}{\pohjaviiva}

%% List of publications

\vsas
\section*{List of Papers}
        \addcontentsline{toc}{section}{List of Papers}
        \addtocontents{toc}{\protect\addvspace{3mm}}

This Thesis consists of an introductory review part,
followed by four research publications:
\medskip

\begin{list}{}{\setlength{\labelsep}{13mm} \setlength{\leftmargin}{15mm}
                 \setlength{\labelwidth}{15mm}}

\item[\zerotxt{I:}] J.~Maalampi, K.~Mursula and R.~Vuopionper\"a,\\
{\em Heavy neutrinos in $e^+e^-$ collisions},\\
Nuclear Physics B {\bf 372} (1992) 23--43.
\bigskip

\item[\zerotxt{II:}] R.~Vuopionper\"a,\\
{\em Heavy Dirac and Majorana neutrino production in $e^+e^-$ collisions},\\
Zeitschrift f\"ur Physik C {\bf 65} (1995) 311--325.

\item[\zerotxt{III:}] R.~Ker\"anen, J.~Pennanen and R.~Vuopionper\"a,\\
{\em How to observe $\nu_\tau N$ interactions at an extremely asymmetric 
$e^+e^-$ collider},\\
Physical Review D {\bf 46} (1992) 4852--4855.
\bigskip

\item[\zerotxt{IV:}] M.~Raidal and R.~Vuopionper\"a,\\
{\em Single top quark production at LEP200},\\
Physics Letters B {\bf 318} (1993) 237--240.\bigskip

\bigskip 

\end{list}

\clearpage

%% Introduction

\vsas
\section{Introduction}   \label{1}

The serious and systematic study of the
structure of matter 
can be said to have
started at the end of the 19th century, and 
it has continued 
ever since with an increasing pace.
Experimental methods for investigating the structure of matter have developed
from tabletop atomic physics measurements, such as the experiments of
Rutherford,
to the international multi-million dollar particle physics 
experiments in huge present day accelerators. At the same
time theories describing the basic laws of Nature have reached very high
level in mathematical rigor. 

The development of physical theories and
the construction of experimental apparatuses have sometimes been rather loosely
connected with each other. However, in high-energy physics, where experiments
nowadays are very large and expensive, it has been realized that one cannot 
construct accelerators and detectors
without having a good understanding of the nature of the phenomena one wants to
investigate.
This is the attitude followed also in this Thesis. We investigate the
phenomenology of electroweak interactions at future high energy colliders and
also propose some new experimental methods to study the constituents of various
particle physics models.

Ever since conjectured by W. Pauli 
in 1930 [\cite{PAULI}], 
neutrinos have played an important role in our
understanding of the basic laws of particle physics. Neutrino was
originally proposed to maintain
the energy conservation law in nuclear beta-decay
[\cite{CONENEL}]. 
One can see, in a sense, a direct route from the neutrino hypothesis to the
present gauge theories of electroweak interactions. The first step along this
route was taken by
Fermi who formulated a field theoretical description for neutrino interactions
[\cite{FERMI}].
In the Fermi theory neutrinos were assumed massless, as was also discussed by
Perrin [\cite{PERRIN}]. A quarter of century later neutrino interactions were
found to be of the $V-A$ form [\cite{VAINTER}], and
the predicted parity violation was discovered in beta decay [\cite{PARITY}].

The existence of the electron neutrino was confirmed in 1956 by Reines and
Cowan [\cite{nuel}], who used an antineutrino beam from a nuclear pile to
detect the reaction $p^+ + \bar{\nu} \rightarrow n + e^+$ followed by the 
reaction
$e^++e^- \rightarrow \gamma + \gamma$ and a $\gamma$ ray emitted from neutron
capture on ${}^{35}{\rm Cd}$ in a large mass scintillator. The fact that
neutrinos emitted in beta decay are left-handed, as the $V-A$ theory predicts,
was confirmed by an angular correction measurement with ${}^{35}{\rm A}$
[\cite{angular}], and also in an experiment which measured the circular
polarisation of the $\gamma$ rays emitted by the exited state of ${}^{152}
{\rm Sm}$ in the electron capture of ${}^{152}{\rm Eu}$ [\cite{exited}].
The existence of s second neutrino, the muon neutrino, was confirmed by Dandy
et al. 
in 1962 [\cite{numu}].

Long before the so-called Standard Model (SM) of electroweak interactions
[\cite{SM}] was formulated in its present form,
it was understood that the Fermi theory is just a
low-energy effective theory [\cite{LOWFER}].
The experimental verification of the SM, whose renormalizability was proven by
't Hooft [\cite{RENOR}], was obtained via the discovery of neutrino induced
neutral current reactions [\cite{NEUCUR}] and finally by the discovery of the
$W^\pm$ and $Z^0$ bosons [\cite{WZ}]. Despite the fact that there are some
unsatisfactory features in SM, by now the model is experimentally very
precisely verified [\cite{PRESI,TP,TOPEVI}].
After the recent discovery of the top quark at TEVATRON [\cite{TOPEVI}], the
only missing constituents of the SM are the Higgs particle and the tau neutrino
($\nu_\tau$). Reactions induced by the tau neutrino have not so far been
directly observed. On the other hand, experiments at the $e^+e^-$ collider
LEP at CERN have shown 
the number of light neutrinos to be three [\cite{PRESI}].
The decay spectrum in some decay channels of the tau-lepton also
indirectly indicates the existence of the tau neutrino [\cite{INDIRT}]. 
The direct discovery of the tau neutrino is a topical issue
[\cite{LHCNEU}], also addressed in this Thesis.

Neutrinos are the only particles in the SM which are sensitive to only one
type of
interaction, the weak force. Therefore neutrinos may play an important role in
revealing possible new physics phenomena, which for other particles
might be shadowed by the effects of electromagnetic and strong interactions.
One important question, where new physics may manifest itself, is neutrino
mass. The ordinary neutrinos are known to be very light ($m_{\nu_e} \lsim {\cal
O}(10)$ eV, $m_{\nu_\mu} \lsim 270$ keV, $m_{\nu_\tau} \lsim 31$ MeV 
[\cite{linuma}]) as
compared with other fermions in the same fermion generation. Finite neutrino
masses would indicate physics beyond the SM, since in the SM neutrinos are
strictly massless.

The precision experiments have confirmed that the gauge interactions
of light neutrinos are quite precisely described by the SM (see e.g. 
[\cite{PHNE}]). 
Apart from the existence of neutrino mass, these experiments leave also many
other questions unanswered. These include the question of
lepton number violation [\cite{LNV}] and the related question of
the nature of neutrino,
i.e. is it a Dirac or a Majorana particle [\cite{MajM}], the neutrino mixing, 
as well as the
possible CP violation in the lepton sector. 
If neutrinos have a non-zero mass 
one must ask whether they are stable particles or not.
For very heavy neutrinos, which have not been discovered so far, 
the nature of gauge couplings is also
an issue one should consider.
The light neutrinos studied in many experiments are known to have left-handed
$V-A$ interactions, but the gauge couplings of the heavy
neutrinos can a priori have a more general 
vector - axial-vector structure as a reflection of
the structure of a possible underlying gauge theory.

From the experimental point of view, the search for small finite neutrino
masses, together with the generation mixing among the light neutrinos,
has so far got the most of the attention 
[\cite{NeMaMi}].
The future accelerators, such as the planned $e^+e^-$ linear colliders
[\cite{Linco1,Linco2}], will offer a good environment for direct searches of
possible heavy neutrinos and for the study of their properties.
This is one of the main topics of this Thesis.
The search of heavy neutrinos 
in proton colliders and in electron-proton
colliders has been considered, e.g., in [\cite{HNLHC}] and [\cite{GrBuHe}].

There 
are plenty of models of electroweak interactions which predict 
heavy neutrinos.
The
simplest of such models is obtained by adding
right-handed heavy neutrinos to the
SM particle spectrum as $SU(2)$ singlets.
If the right-handed singlets are not present,
but neutrinos have a mass, then they are necessarily Majorana particles.
This is the situation in the SM as well as in the grand unified theories
based on $SU(5)$ symmetry [\cite{SU5}].
In the left-right symmetric model (LR-model)  [\cite{LR}], whose
gauge symmetry is $SU(2)_R\otimes SU(2)_L\otimes U(1)_{B-L}$, one has heavy
Majorana neutrinos and in some versions of the model also heavy Dirac neutrinos
[{\cite{MRHN}]. The existence of these heavy states is related to the so-called
see-saw mechanism [\cite{seesaw}], in terms of which one can explain
the tiny mass of the left-handed neutrinos in a very attractive way.
The LR model is naturally embedded in $SO(10)$-based grand unified theories 
(GUT) [\cite{SO10}],
which have the nice feature that all the fermions of a given family, including
the right-handed neutrinos, are assigned into an irreducible anomaly free
representation.
Among many other models that predict heavy neutrinos one should mention the
superstring inspired $E_6$ models [\cite{E6}].

This Thesis is organized as follows. First we give a brief summary of the
appended original research Papers. 
In Section \ref{2} we describe 
the central elements of the Standard Model,
in particular its particle spectrum from the point of view of neutrino physics.
Some aspects of the SM phenomenology at the $e^+e^-$ colliders are also
briefly discussed in this Section.
Section  \ref{3} deals with extended gauge models.
The physics of heavy neutrinos in the left-right symmetric electroweak model
is studied in more detail. 
In Section \ref{4} we present our conclusions.

\bigskip

\newpage

\vsas
\subsection*{Summary of the Original Papers}   \label{1.2}
        \addcontentsline{toc}{subsection}{Summary of the Original Papers}
        \addtocontents{toc}{\protect\addvspace{3mm}}

\medskip

\paragraph{Paper~I:
Heavy neutrinos in $e^+e^-$ collisions.}
In this Paper the production and the subsequent decay of heavy neutrinos
in $e^+e^-$ collision are systematically investigated.
The purpose is to
study the possibility of
distinguishing heavy Dirac and
Majorana neutrinos ($N$) from each other in electron-positron collisions.
In the previous studies [\cite{heoth}] the main attention has
been paid on the nature of the {\em light} neutrinos.
A new experimentally easy method for separating the Dirac and
Majorana cases
is presented. It is based on 
equal-sign lepton pair correlations in the decay of $N\bar{N}$ system. This
method would
work well even with low statistics.
General formulas for the pair and single heavy neutrino 
production cross sections and the decay widths of heavy neutrinos are derived. 
Also other experimental methods for distinguishing Dirac and Majorana cases
are discussed on the basis of these formulas.
The experimental signals of the heavy neutrino production are shown to be
clean and almost background free,
especially in the $\nu N$ production channel where
monojets and otherwise strongly unbalanced events will appear.
\medskip

\paragraph{Paper~II:
Heavy Dirac and Majorana neutrino production in $e^+e^-$ collisions.}
This Paper broadens the analysis of the Paper I by considering the production 
of heavy neutrinos within a more general theoretical framework.
The effects of
non-standard
Higgs bosons to the heavy neutrino production 
in electron-positron collisions are investigated systematically.
The analytical differential cross section formulas
are derived using the most general Lagrangian with complex
lepton and boson couplings. 
For the total cross sections 
the numerically integrated results are shown.
All the interference terms are carefully studied
and they are found to be fairly large in a wide range of the production
spectrum. 
Using a simple version of the
left-right symmetric model, specified in the Paper, the
heavy neutrino signal is found to be clearly visible at a $500$ GeV linear
$e^+e^-$ collider  with a luminosity around
${\cal L}_{\rm year} = 10\; {\rm fb}^{-1}$. Even the detection
of the triplet higgs $\Delta^{\pm}$ with a mass around $1$ TeV is found 
marginally possible if a second phase of the $e^+e^-$ collider with a higher
collision energy ($\sqrt{s} = 1.0 - 2.0$ TeV) and a higher luminosity 
(${\cal L}_{\rm year} \approx 100\; {\rm fb}^{-1}$) is constructed after the
preliminary $0.5$ TeV phase.

\paragraph{Paper~III:
How to observe $\nu_\tau N$ interactions at an extremely asymmetric 
$e^+e^-$ collider.}
The weakly interacting partner of the tau lepton, the tau neutrino, has not
been directly observed, although its existence is indirectly verified through
$\tau$ decays [\cite{INDIRT}].
Also results of the LEP and SLC have confirmed 
that in addition to the electron neutrino and the muon neutrino a third light
neutrino species exists.
In this Paper a new and novel idea for an accelerator and detector
concept for a direct tau neutrino discovery is proposed.
It is shown that by using a
very asymmetric ring-linac type [\cite{rili}] 
electron and positron collider one could produce a well-focused
and monoenergetic tau neutrino beam. A robust layout for a whole accelerator
and detector setup is presented and the most critical technical aspects
are studied. Using Monte Carlo simulation
a method to separate the signal from background is presented.
The luminosity required for the tau neutrino
detection is found to be in the range ${\cal L} \approx 10^{34}\;
{\rm cm}^{-2}{\rm s}^{-1}$, i.e. comparable with the luminosity of the proposed
electron-positron colliders [\cite{Linco1,Linco2}].
\medskip

\paragraph{Paper~IV:
Single top quark production at LEP200.}
In this Paper the
prospects for detecting the top quark at the LEP200 collider are investigated.
It is argued that 
it would be marginally possible
to observe the single top production at LEP200.
The cross section of
the reaction $e^+e^-\rightarrow t\bar{b} 
e^-\bar{\nu}_e\; (
\bar{t}b e^+\nu_e)$, is estimated by integrating the analytical eight
dimensional differential cross sections by using Monte Carlo integration
method. Later more detailed numerical and analytical studies by other authors
[\cite{topot}]
have shown, however, that our results overestimate the production rate and
our conclusions may therefore be too optimistic as far as LEP200 is
concerned. For higher energy $e^+e^-$ collisions the process considered will be
an important source of the top quarks.
\medskip

\newpage

\vsas
\section{The Standard Model}     \label{2}

The Weinberg-Salam model
[\cite{SM}] is the most
successful theory for the electroweak interactions of quarks and leptons. It is
a renormalizable field theory based on the $SU(2)_L\otimes U(1)_Y$ gauge
symmetry spontaneously broken to the residual $U(1)_{em}$ symmetry
of electromagnetic interactions.
To implement the spontaneous breaking
one has to  include scalar particles, Higgs bosons, to the particle spectrum 
of the model.
The spontaneous symmetry breaking generates masses of the weak bosons $W^\pm$
and $Z^0$ through the so-called Higgs mechanism [\cite{Higgs}], and at the
same time masses of quarks and leptons via the Yukawa couplings 
of scalars and fermions.
The strong interactions of quarks described by the quantum chromodynamics (QCD)
are included to the model by adding an $SU(3)_C$ gauge symmetry to the
Weinberg-Salam symmetry, which then 
together form
the Standard Model (SM).

The particle spectrum of the SM consists of scalar Higgs
particles, vector bosons and spin-${1}/{2}$ fermions. The fermions,
i.e. leptons and quarks,
are grouped into three fermion generations.
Under the $SU(2)_L\otimes U(1)_Y\otimes SU(3)_C$ gauge symmetry each fermion
generation ($k=\samepage 1,2,3$) has the following field contents 
(the color indices of the quark fields are suppressed):
\begin{equation}
\begin{array}{lclclclclclcl}
L_k&=&\left(\begin{array}{c} \nu_{\ell_k}\\ \ell_k\end{array}\right)_L
&\!\sim\!&
(2,-1,1),&\!\!\!&\ell_{kR} &\sim& (1,-2,1),& & & &\vspace{2.0mm}\\
Q_{kL}&=&\left(\begin{array}{c} u_k\\ d_k\end{array}\right)_L&\!\sim\!&
(2,\frac{1}{3},3),&\!\!\!&u_{kR}&\sim&(1,\frac{4}{3},3),&\!\!\!&d_{kR}
&\!\sim\!& (1,-\frac{2}{3},3).
\end{array}\label{SMPSp}
\end{equation}
The right-handed neutrinos, not present in the minimal version of the SM,
would be totally inert with respect to gauge interactions, i.e. they would 
transform as $\nu_{kR}\sim(1,0,1)$.

The scalar sector of the SM
consists of a Higgs doublet
\begin{equation}
\phi \equiv \left(\begin{array}{c} \phi^+ \\ \phi^0\end{array} \right)
\sim (2,1,1),
\end{equation}
where the neutral component acquires vacuum expectation value,
thereby breaking the electroweak symmetry.
This is the minimal
case, but in general the Higgs sector may be more complicated.
For example, in the minimal supersymmetric version of the SM
(MSSM) [\cite{SUSY}] there should exist two scalar doublets, at least. One can
also add different scalar representations (e.g. $SU(2)_L$-triplets) to the 
scalar particle spectrum, as will be described later on.

The Lagrangian of the SM is of the form
\begin{equation}
{\cal L} =
{\cal L}^{matter}_{kin} + {\cal L}^{gauge}_{kin} 
+ {\cal L}_Y - V(\phi).
\end{equation}
The kinetic terms are given by
\begin{eqnarray}
{\cal L}^{matter}_{kin} &=& \left|\left(\partial_\mu-
\dot{\imath}\,\frac{g}{2}\,\overline{\tau}\cdot \overline{W}_\mu-
\dot{\imath}\,\frac{g'}{2}
B_\mu\right)\!\phi\,\right|^2\nonumber\\
& &+\dot{\imath}\,\bar{Q}_{kL}\gamma^\mu\left(\partial_\mu
-\dot{\imath}\,\frac{g}{2}\,\overline{\tau}\cdot \overline{W}_\mu -
\dot{\imath}\,\frac{g'}{6}
B_\mu\right)Q_{kL}\nonumber\\
& &+ \dot{\imath}\,\bar{u}_{kR}\gamma^\mu 
\left(\partial_\mu-\dot{\imath}\,\frac{2g'}{3}
B_\mu\right)u_{kR}+\dot{\imath}\,
\bar{d}_{kR}\gamma^\mu \left(\partial_\mu+\dot{\imath}\,\frac{g'}{3}
B_\mu\right)d_{kR}\label{SMkin}\\
& & +\dot{\imath}\,\bar{L}_k\gamma^\mu\left(\partial_\mu
-\dot{\imath}\,\frac{g}{2}\,\overline{\tau}\cdot \overline{W}_\mu +
\dot{\imath}\,\frac{g'}{2}
B_\mu\right)L_k\nonumber\\
& &+\dot{\imath}\,\bar{\ell}_{kR}\gamma^\mu \left(\partial_\mu+\dot{\imath}\,g'
B_\mu\right)\ell_{kR}
+\dot{\imath}\,\bar{\nu}_{kR}\gamma^\mu \partial_\mu\nu_{kR}\nonumber
\end{eqnarray}
and
\begin{equation}
{\cal L}^{Gauge}_{Kin} = -\frac{1}{4}\,\overline{W}_{\mu\nu}\cdot
\overline{W}^{\mu\nu}-
\frac{1}{4}\,B_{\mu\nu}B^{\mu\nu}\, ,
\end{equation}
where
\begin{equation}
\overline{W}_{\mu\nu} = \partial_\mu\overline{W}_\nu
-\partial_\nu\overline{W}_\mu + g\,\overline{W}_\mu\times \overline{W}_\nu\, ,
\end{equation}
and $\overline{W}_\mu$ stands for ($W_{1\mu}$, $W_{2\mu}$, $W_{3\mu}$).
We have included in the Lagrangian (\ref{SMkin}), and in what follows, also
the right-handed neutrinos to serve our later considerations.

The Yukawa terms 
describing interactions of the quarks and leptons with the Higgs scalars
are given by the Lagrangian
\begin{equation}
-{\cal L}_Y = \left(h_d\right)_{kj}\,\bar{Q}_{kL}\phi d_{jR} + 
\left(h_u\right)_{kj}\,\bar{Q}_{kL}\tilde{\phi} u_{jR} + 
\left(h_\ell\right)_{kj}\,\bar{L}_k\phi \ell_{jR} +
\left(h_\nu\right)_{kj}\,\bar{L}_k\tilde{\phi}\nu_{jR} + h.c. \, ,
\label{Ykaw}
\end{equation}
where $h_d$, $h_u$, $h_\ell$ and $h_\nu$ are dimensionless Yukawa coupling
matrices and the charge conjugate of the scalar field is defined as
$
\tilde{\phi} = \dot{\imath}\tau_2\phi^* 
\sim (2,-1,1)
$.
The last term of (\ref{Ykaw}) gives rise to
neutrino masses via the spontaneous symmetry breaking.
In the minimal SM it does not exist, and the neutrinos are massless.

The Higgs potential is chosen to be
\begin{equation}
V(\phi) = -\mu^2(\phi^\dagger\phi) + \frac{\lambda}{4}
(\phi^\dagger\phi)^2\,,\;\;\;
\mu^2 >0.
\end{equation}
The minimum of this potential corresponds to the vacuum expectation
value
\begin{equation}
\left<\phi\right> =\frac{1}{\sqrt{2}}
\left(\begin{array}{c}0\\ v\end{array}\right)
\end{equation}
of the scalar doublet, where 
$v=2\mu / \sqrt{\lambda}$. 
The vacuum breaks the electroweak symmetry $SU(2)_L\otimes U(1)_Y$, except
the  $U(1)_{em}$ subsymmetry associated with the electromagnetism.
The gauge bosons, $W^\pm$ and $Z^0$,
corresponding to the broken generators acquire mass [\cite{SM}], while the
gauge boson, photon, 
corresponding to the unbroken generator of $U(1)_{em}$ remains massless. 
In other words,
the SM symmetry $SU(2)_L\otimes U(1)_Y\otimes SU(3)_C$ is spontaneously broken
down to the $U(1)_{em}\otimes SU(3)_C$ residual gauge symmetry.
The masses of the weak bosons $W^\pm$ and $Z^0$ are
\begin{equation}
M_W=\frac{gv}{2}\;,\; 
\;\;\; M_Z=\frac{v\sqrt{g^2+{g'}^2}}{2}=\frac{M_W}{\cos\theta_W}\, ,
\label{SMBosMa}
\end{equation}
where $\theta_W$ is the so-called weak mixing angle.

In terms of the physical gauge fields the interaction Lagrangian is given by
\begin{eqnarray}
{\cal L}_{NC}&=& -e J^\mu_{em}A_\mu - GJ^\mu_{NC} Z_\mu\;\; =\;\;
-e\,J^\mu_{em}A_\mu - G\left(J^\mu_3-\sin^2\theta_WJ^\mu_{em}\right) Z_\mu\, ,
\label{LNC}\\
{\cal L}_{CC}&=& -\frac{g}{2\sqrt{2}}\,J^\mu_{CC}W^+_\mu + h.c.\, ,
\label{LCC}
\end{eqnarray}
where 
\begin{eqnarray}
G^2 &=& g^2+{g'}^2\, ,\\
J^\mu_{NC} & = & J^\mu_3-\sin^2\theta_WJ^\mu_{em}\, , \label{NCLA}\\
J^\mu_{CC} & = & 2\left[\bar{\nu}_{kL}\gamma^\mu \ell_{kL}
+\bar{u}_{kL}\gamma^\mu d_{kL}\right]\, , \label{CCLL}\\
J^\mu_{em} & = & -\bar{\ell}\gamma^\mu \ell
+\frac{2}{3}\bar{u}\gamma^\mu u
-\frac{1}{3}\bar{d}\gamma^\mu d\, , \label{NCEM}\\
J^\mu_{3} & = &  -\frac{1}{2}\bar{\ell}_{L}\gamma^\mu\ell_{L}
+\frac{1}{2}\bar{\nu}_{L}\gamma^\mu \nu_{L}
-\frac{1}{2}\bar{u}_{L}\gamma^\mu u_{L}
+\frac{1}{2}\bar{d}_{L}\gamma^\mu d_{L}\, .\label{NCZZ}
\end{eqnarray}
The weak mixing angle $\theta_W$ 
has the experimentally measured value $\sin^2\theta^{LEP}_W = 0.2319\pm
0.0005\pm0.0002$ [\cite{SINLEP}]. 

The quark and lepton fields appearing in the above interaction Lagrangian,
so-called weak eigenstates, do not in general correspond to the physical
propagating particle states with a definite mass. The mass eigenstates are
obtained by diagonalizing the Yukawa Lagrangian (\ref{Ykaw}) by suitable 
unitary transformations. For example, the physical up-type quarks $u',c',t'$
are given in terms of the interaction eigenstates as follows:
\begin{eqnarray}
\left(\begin{array}{c} u'\vspace{-5mm}\\ c'\vspace{-5mm}\\ t' 
\end{array}\right)_L & = & 
U_u \left(\begin{array}{c} u\vspace{-5mm}\\ c\vspace{-5mm}\\ t 
\end{array}\right)_L\, ,\nonumber\\
 & &\\
\left(\begin{array}{c} u'\vspace{-5mm}\\ c'\vspace{-5mm}\\ t' 
\end{array}\right)_R & = & 
V_u \left(\begin{array}{c} u\vspace{-5mm}\\ c\vspace{-5mm}\\ t 
\end{array}\right)_R\,,\nonumber
\end{eqnarray}
where $U^\dagger_u U_u = V^\dagger_u V_u = 1$.

Only the form of the charged current part of the gauge Lagrangian (\ref{LCC})
is changed when one moves from the interaction basis of the fermion fields
into the mass basis.
It reads now
\begin{equation}
J^\mu_{CC} = 2\left[\bar{\nu}_{jL}' (U^{CKM}_\ell)_{jk}
\gamma^\mu \ell_{kL}'
+\bar{u}_{jL}'(U^{CKM}_q)_{jk}\gamma^\mu d_{kL}'
\right]\, ,\label{CCLA}
\end{equation} 
where the complex 
unitary
Cabibbo-Kobayashi-Maskawa mixing 
matrices are
$U^{CKM}_\ell = U^\dagger_\nu U_l$ and $
U^{CKM}_q = U^\dagger_uU_d$.
The $U^{CKM}_q$ matrix, except for the $U_{tx}$ elements, is experimentally  
quite well known [\cite{UCKMVA}]. Even the  $U_{tx}$ elements can be fairly
well limited on the grounds of the unitarity of $U^{CKM}_q$.
In contrast, the elements of the lepton mixing matrix $U^{CKM}_\ell$ are
almost unknown. The results
of neurinoless beta decay experiments give only weak constraints on 
$U^{CKM}_\ell$ [\cite{DBDUli}].

Obviously the mixing formalism is valid for any number $n_f$ of fermion
generations, and the number of the quark and the lepton generations may also
differ in a general case.
The $n_f\times n_f$ 
unitary matrices can be parameterized with $n_f^2$ real parameters, i.e.  
with $\frac{1}{2}n_f(n_f-1)$ rotational angles and $\frac{1}{2}n_f(n_f+1)$
phases. 
In the case of fermion mixing, however,  $2n_f-1$ phases can be absorbed by
redefining the physical fermion fields and therefore the CKM
matrices can be parameterized 
in terms of
$\frac{1}{2}n_f(n_f-1)$ rotational angles and  $\frac{1}{2}(n_f-1)(n_f-2)$
phases [\cite{PARA}]. In the case of three generations this means three
rotation angles and one phase. 

The non-diagonal nature of the charged currents permits flavor changing
reactions, such as $K^+ \rightarrow \mu^+\nu_\mu$, where one has the transition
$\bar{s}\rightarrow \bar{d}$ [\cite{FCCC}], and the CP violating reactions,
such as $K^0_L\rightarrow \pi^+\pi^-$
[\cite{CPVIO}], arise because of the complex phase in the mixing matrix.
Flavor changing neutral current
and CP violating reactions have been observed in the quark sector
[\cite{OFCCP}], but never
in the lepton sector, where the very small neutrino mass
forces the corresponding decay widths and cross sections to be negligible.
If neutrinos do not have a mass there is neither flavor changing neutral
currents nor CP violating reactions in the lepton sector because the mass
eigenstates of the leptons are in this case the same as the current 
eigenstates.

\vsas
\subsection{Neutrino Masses in $SU(2)_L\otimes U(1)_Y$ Model} \label{2.1}

If 
the right-handed 
neutrino fields 
exist, the 
mass generation
mechanism described above for quarks and charged leptons would 
also give rise to massive neutrinos in the SM.
Such massive neutrinos will be
Dirac particles.
In this case the physics of the lepton sector would be very similar to that of
quarks. In particular,
the lepton generation mixing is 
analogous to the quark generation mixing,
and the mixing matrix could be presented in the general form
\begin{equation}
U^{CKM}_{q\,(\ell)} = \left(\begin{array}{lcr}
c_{\theta_2}c_{\theta_1} & c_{\theta_2}s_{\theta_1} & s_{\theta_2}\\
\!\!\!{\scriptstyle -} c_{\theta_3}s_{\theta_1}e^{\dot{\imath}\delta}
{\scriptstyle -} s_{\theta_3}s_{\theta_2}c_{\theta_1}& c_{\theta_3}
c_{\theta_1} e^{\dot{\imath}\delta}{\scriptstyle -}s_{\theta_3}s_{\theta_2}
s_{\theta_1}& s_{\theta_3}c_{\theta_2}\\
s_{\theta_3}s_{\theta_1}e^{\dot{\imath}\delta}{\scriptstyle -}c_{\theta_3}
s_{\theta_2}c_{\theta_1}& {\scriptstyle -} s_{\theta_3}c_{\theta_1}
e^{\dot{\imath}\delta}{\scriptstyle -} c_{\theta_3}s_{\theta_2}s_{\theta_1}
& c_{\theta_3}c_{\theta_2}\\
\end{array}\right)\, . \label{CKMMa}
\end{equation}
As a result of the mixing the flavor changing charged lepton currents will
hence be allowed.

The mass mixing would also make possible the neutrino
oscillations [\cite{NUOscV,NUOscM}], i.e.
\begin{equation}
\nu_{\ell} \leftrightarrow \nu_{\ell'} \;\;\;\; {\rm and} \;\;\;\;
\nu^c_\ell \leftrightarrow \nu^c_{\ell'}\, \;\;\;\; \left(\ell\neq \ell'
\right)\, .\label{NeuO}
\end{equation}
On the other hand, neutrino-antineutrino oscillations  $\nu_{\ell} 
\leftrightarrow\nu^c_{\ell'}$ are forbidden. This 
is because the total lepton number $L = \sum_{i=1}^n L_i$ 
is conserved by the Dirac mass terms.

If there are no right-handed neutrinos in the theory,
the above
mass generation mechanism does not work. The only allowed mass
term in this case is the so-called Majorana mass term
\begin{equation}
{\cal L}_\nu^M = \left(m^M_\nu\right)_{ij}\nu^T_{iL}C\nu_{jL}
+{\rm h.c.} \equiv  \left(m^M_\nu\right)_{ij}\bar{\nu}_{iR}^c\nu_{jL}
+{\rm h.c.}\,,\label{MajMas}
\end{equation}
where $C$ is the charge conjugation matrix and $\bar{\nu}_R^c = \nu^T_L C$.
As such this would violate the gauge invariance and is therefore not present
in the original Lagrangian, but it may enter as a result of a
spontaneous symmetry breaking involving an $SU(2)_L$ triplet scalar multiplet
$\Delta \sim (3,2,1)$. The corresponding
Yukawa term is
\begin{eqnarray}
{\cal L}_Y'' &=& \left(h_\ell''\right)_{ij}\left(\nu_{iL}^T\:
\ell_{iL}^T\right)
C\left(\dot{\imath}\sigma_2\right)\vec{\sigma}\left(\begin{array}{c}
\nu_{jL}\\ \ell_{jL} \end{array}\right) \cdot \vec{\Delta} +{\rm h.c.}\\
             &=&\left(h_\ell''\right)_{ij}\left(
-\bar{\ell}_{iR}^c\: \bar{\nu}_{iR}^c\ \right)\left(\begin{array}{ll}
\Delta_3 & \Delta_1- \dot{\imath} \Delta_2\\
\Delta_1+ \dot{\imath} \Delta_2 &-\Delta_3\end{array}\right)\left(
\begin{array}{c}\nu_{jL}\\ \ell_{jL} \end{array}\right)+{\rm h.c.}\\
             &=&\left(h_\ell''\right)_{ij} \left[ 
-\frac{1}{\sqrt{2}}\left(\bar{\ell}_{iR}^c \nu_{jL}+
\bar{\nu}_{iR}^c\ell_{jL}\right)\Delta^+-
\bar{\ell}_{iR}^c \ell_{jL}\Delta^{++} +\bar{\nu}_{iR}^c\nu_{jL}\Delta^0
\right] + {\rm h.c.}\\
             &=&-\left(h'_{+}\right)_{ij}\bar{\ell}_{iR}^c
\nu_{jL}\Delta^+ -\left(h''_{+}\right)_{ij}
\bar{\nu}_{iR}^c\ell_{jL} \Delta^+\nonumber\\
&& -\left(h'_{++}\right)_{ij}\bar{\ell}_{iR}^c \ell_{jL}\Delta^{++}
+\left(h'_0\right)_{ij}\bar{\nu}_{iR}^c \nu_{jL}\Delta^0
 + {\rm h.c.}\, ,\;\;\;\;\;\;\; {}^{}\label{last}
\end{eqnarray}
where $\Delta^0=\Delta_1+ \dot{\imath} \Delta_2$, $
\Delta^+=\sqrt{2}\,\Delta_3$ and
$\Delta^{++}=\Delta_1- \dot{\imath} \Delta_2$. The last equation (\ref{last})
is the most general form of this kind of Yukawa coupling terms.
The 
Majorana mass term (\ref{MajMas}) arises when the vacuum expectation value
$\left<\Delta^0\right>=v_L\neq 0$, and the 
Majorana neutrino mass matrix of the Standard Model 
appearing in (\ref{MajMas}) is thus
\begin{equation}
\left(m_\nu^M\right)_{ij} = \left(h'_0\right)_{ij}\left<\Delta^0\right>
\,.
\end{equation}

It should be emphasised that the spontaneous breaking of the $SU(2)_L\otimes
U(1)_Y$ symmetry does not require the existence of a triplet Higgs $\Delta$.
Actually, the experimentally measured value of the parameter $\rho_0 = 
\left(M_W/\cos\theta_W M_Z\right)^2 = 1.0004\pm0.0022\pm0.002$
[\cite{rhopar}] indicates that the vacuum
expectation value of the left-handed triplet Higgs field should
be much smaller than that of the doublet scalar field $\phi$, i.e. $\langle
\Delta_L\rangle \lsim {\cal O}(10)$ GeV $\ll \langle\phi\rangle \equiv
v/\sqrt{2} \simeq 175$ GeV.
Hence, the existence of the triplet $\Delta_L$ is not strongly motivated, and
one can say 
that neutrinos are most naturally massless in the $SU(2)_L\otimes U(1)_Y$ 
model.

The generation mixing in the case of Majorana neutrinos 
differs from the quark generation mixing in many respects. Due to Fermi
statistics the mass matrix must be symmetric $M\equiv \left(m^M_\nu\right) = 
\left(m^M_\nu\right)^T$. One can diagonalize such a symmetric and complex mass 
matrix by the following transformation:
\begin{equation}
m = {\rm diag}(m_1,\ldots,m_n) = U^TMU\, ,\label{dimf}
\end{equation}
where 
$m_i \geq 0$ and $U$ is unitary matrix. In contrast to the Dirac mass case,
only one unitary matrix is needed to diagonalize a Majorana mass matrix. This
is of course a consequence of the fact that the Majorana mass term involve
just the left-handed fields.
The mass Lagrangian can be rewritten in the form
\begin{eqnarray}
{\cal L}^M_\nu &=& 
-\frac{1}{2}\bar{\chi} m \chi=-\frac{1}{2}\sum_{i=1}^n m_i\bar{\chi}_i\chi_i\\
&=&-\frac{1}{2}\left(\bar{n}^c_Rmn_L-\bar{n}_Lmn^c_R\right)
=-\frac{1}{2}\left(n^T_LCmn_L-{n^c_R}^TCn^c_R\right)\,,
\end{eqnarray}
where $n_L = U^\dagger \nu_L$, $n^c_R = C\bar{n}_L^T=U^T \nu^c_R$ and
\begin{equation}
\chi = n_L + n^c_R = \left(\begin{array}{c} 
\chi_1\vspace{-5mm}\\ \chi_2\vspace{-5mm}\\ \vdots\vspace{-5mm}\\ \chi_n
\end{array} \right)\, .
\end{equation}
The chiral composition of the mass eigenstates $\chi_i$ is the following:
\begin{equation}
\chi_i = \chi_{iL} + \chi_{iR} = n_{iL} + n^c_{iR} = \sum_{j=1}^n \left[ 
\left(U^\dagger\right)_{ij}\nu_{jL}+\left(U^T\right)_{ij}\nu^c_{jR}\right]\, .
\end{equation}
The states obey $\chi_i = \chi^c_i = C\bar{\chi}^T_i$, i.e. they are Majorana
particles.

The current states are given in terms of mass eigenstates via the formula
\begin{equation}
\nu_{jL} = \sum_{i=1}^n \left(U\right)_{ji}\chi_{iL}\label{diml}.
\end{equation}
The functional form of the charged current 
Lagrangian for the mass eigenstates $\chi_i$ is
exactly the same as in
the Dirac case (eq. (\ref{CCLA})), except that the $\nu'_i$
fields are replaced by the $\chi_i$ fields. The neutral current
term (see eqs. (\ref{NCLA}), (\ref{NCEM}) and (\ref{NCZZ})) remains flavor
diagonal, but the physical $\nu'_i$ fields are replaced by $\chi_i$ fields.

In the Majorana case the total lepton number is an invariant in the charged and
neutral current gauge interactions but the Majorana mass term breaks the lepton
number by two units.
So, in the case of the Majorana
neutrinos both the neutrino-neutrino oscillations and
the neutrino-antineutrino oscillations are possible. 

\vsas
\subsection{Standard Model Phenomenology
at $e^-e^+$ Collisions}   \label{2.2}

Some of the features of the SM
have not been directly confirmed by measurements yet. 
Apart from the Higgs particle, the only undiscovered constituent of the SM
particle spectrum is the neutral partner of the $\tau$ lepton, the 
tau neutrino $\nu_\tau$.
The heaviest quark, the $t$-quark, 
another member of third fermion
generation, has recently been discovered at Tevatron [\cite{TOPEVI}].
The mass and the other characteristics of the top quark are still quite poorly
known. 

The experimental tests of the SM predictions have by now reached the level of 
a few per mils [\cite{LEPtest}]. While most of the recent particle discoveries
have been made by using
hadron colliders, e.g. the discovery of the $W$ and $Z$ bosons at SPS
and the top quark at Tevatron, the precision tests are mainly due to lepton
collision experiments.
Due to the 
overwhelming background problems
at future hadron experiments, one expects the role of lepton machines
to become more important in searching new physics in the future.
The discovery of the SM Higgs particle, not considered in this Thesis, will
obviously be one of the main goals of the future $e^+e^-$ colliders.

In Papers III and IV we study possibilities to 
investigate the tau neutrino and the top quark in experiments 
with lepton beams.
In Paper III a new idea of using a very asymmetric $e^+e^-$ collider
to create a monoenergetic tau neutrino beam of high intensity is proposed.
Asymmetric $e^+e^-$ collisions are achieved by using electron and positron
beams with unequal beam energies.
Adjusting the beam energies properly one can produce $\nu_\tau\bar{\nu}_\tau$
and $\tau^+\tau^-$ pairs in a highly boosted CM frame.
Thus, most of the tau neutrinos are boosted to very narrow cone in the
laboratory frame. 
The energy and the production angle of neutrinos are correlated. 
For observing the tau neutrino via the reactions it induces in matter, we
propose a long coarsely instrumented iron spectrometer.

The feasibility of the proposed experimental setup is studied using 
computer simulations which took into account the physical processes and the
experimental constraints.
The signal is found to be
marginally observable if the high-energy beam parameters, e.g. the
beam size and beam intensity at the interaction point, designed for the
future $e^+e^-$ colliders, are used.
On the other hand, the signal is found to be almost background free
due to the experimental setup which is designed in such a way that only
neutrinos can enter the detector.
If the luminosity of the accelerator and/or the volume of the detector
are further increased, the expected signal will, of course, be enhanced.
For detecting the tau neutrino also other methods have been considered in
the literature, such as beam dumb experiments [\cite{Telnov}] and the proposed
tau neutrino experiment at the LHC [\cite{LHCNEU}]. As compared with these, the
arrangement we propose has the virtue of giving an almost monoenergetic
neutrino beam and it also has a much better signal to background ratio.

The main focus of Paper IV is the top quark.
The work was done prior to the discovery of the top quark at Tevatron
[\cite{TOPEVI}]. 
We investigated the possibilities discovering and studying the
top quark at the upgraded LEP200 via
the single top production
reaction $e^+e^-\rightarrow t\bar{b} e^-\bar{\nu}_e$
($\bar{t}b e^+\nu_e$). 
The total number of tree level amplitudes for this reaction is 
thirty, but
most of them give a negligible contribution. We included in our
calculations 
eleven graphs.
Due to the very small Yukawa coupling between the Higgs boson and the electron,
the processes involving a neutral Higgs are negligible and are not taken into
account.
Due to the high mass of the $Z^0$ boson the phase space is rather limited
and thus the production rates are strongly reduced in the $Z^0$ exchange
reactions compared with the dominant photon processes
(see ref. [\cite{LEP200top}]).

The results of our calculations made us to conclude that it
would be marginally possible to observe the top quark at the upgraded LEP200.
Later and more detailed studies by other authors [\cite{topot}] have
shown, however, that our production rate estimates are inaccurate and
for that reason too optimistic.
The origin of the inaccuracy of our analysis can be traced back
to large cancellations between different individual processes which
caused undetectable numerical errors in our Monte Carlo integration program. 

Thanks to the efforts of many authors [\cite{topot}]
the single top production process at $e^+e^-$ collisions is now totally
under control and well understood. 
While not important at LEP200, this process will play an important role as a 
$t$-quark source in the future high-energy $e^+e^-$ colliders [\cite{MELE}].

\newpage

\vsas
\section{Extended Gauge Models}   \label{3}

In spite of its theoretical appeal and excellent phenomenological success, the
Standard Model has some unsatisfactory features. 
It suffers from
the naturalness problem  [\cite{Natu}] and the hierarchy problem
[\cite{Hier}] and, furthermore, it does not give any explanation for the
the maximal parity violation [\cite{Parity}] or for the origin of CP violation
[\cite{CPVio}].
There is also a great number of free parameters, $19$ altogether, which
one can fix only by experimental measurements.
Also, the SM 
does not explain the family structure of the fermions. 

Some of these shortcomings seem to point towards a further unification of
electroweak and strong interactions in so-called grand unified theories
(GUT's), some towards supersymmetry [\cite{SUsy}], compositeness 
[\cite{COMPO}] or so-called technicolor schemes [\cite{TC}].
The most studied GUT models are the $SU(5)$ and $SO(10)$ theories,
and due to the LEP results [\cite{LEPsu5}] the supersymmetric versions of
these models seem to be favored at the moment. The preonic composite models 
have been recently studied extensively in the literature [\cite{preonic}]. The
technicolor models and similar composite models 
[\cite{COM_TC}] are somewhat
disfavored by current experimental results (see e.g. [\cite{SR_EW}]).

In this Thesis we will mainly concentrate on the gauge models
of electroweak interactions
based on a larger gauge group than the $SU(2)_L\otimes U(1)_Y$
of the SM. In general, the only requirements we impose on such an extended
gauge model is that it should contain all the SM particles and that its gauge
symmetry must eventually break down to the SM gauge group.
In general,
we are also interested in unified theories which, unlike the 
$SU(3)_C\otimes SU(2)_L\otimes U(1)_Y$ symmetric SM, does unify all the forces 
within one gauge group.
All this can be accomplished using higher symmetry groups which contain the
SM gauge group as a 
subgroup. 

The breaking of the grand gauge symmetry ($G_U$) to the SM symmetry proceeds
in general through several intermediate stages:
\begin{equation}
G_U \stackrel{M_U}{\rightarrow} G_{I_1}
\stackrel{M_{I_1}}{\longrightarrow} \ldots 
\stackrel{M_{I_{n-1}}}{\longrightarrow} G_{I_n} 
\stackrel{M_{I_n}}
{\rightarrow} SU(3)_C\otimes SU(2)_L\otimes U(1)_Y\,,\label{Break}
\end{equation}
where the intermediate breakings 
occur at energy scales $M_U$ and $M_{I_1},\ldots, M_{I_n}$. 

In this Thesis particular attention is paid to
such extended gauge models which
at some stage of the breaking chain contain an
$SU(2)_R\otimes SU(2)_L\otimes U(1)_{B-L}$ gauge symmetry, the so-called
left-right electroweak symmetry [\cite{LR}].
Otherwise the considered extended gauge models are only restricted by 
the experimental results, i.e. new interactions should not interfere with the 
successful predictions of the SM, and 
new particles predicted by the models must be heavier than 
the corresponding experimental lower bounds on masses.
The new particles include new vector bosons
corresponding to the new generators of the extended gauge symmetry, in general
also new fermions, such as, heavier replica of known fermions, heavy
neutrinos and, e.g., so-called mirror fermions [\cite{Mirror}], as well as new 
Higgs scalars.

\vsas
\subsection{Extended Gauge Models for Massive Neutrinos}   \label{3.1}

Possible models for massive neutrinos can be divided into two 
categories, according to whether the neutrinos are
Majorana or Dirac particles.
In the
category of Majorana neutrinos there are models where
the light neutrino masses are generated via
the see-saw mechanism [\cite{seesaw}], as well as models 
where the neutrino masses are induced by radiative corrections [\cite{MMbyRC}].

The see-saw mechanism is the simplest way to understand the
lightness of the 
left-handed neutrinos, and it is realized in many extended models.
A central ingredient of this mechanism is a Majorana mass term of
the right-handed neutrinos, which is much larger than any mass scale of the SM.
As for all mass parameters, the origin of this mass should be connected to some
symmetry breaking taking place in the theory.
The models 
with this property include the local left-right symmetric model,
local $SO(10)$ based models [\cite{SO10}],
models with local or global horizontal symmetry [\cite{Horis1,Horis2}] and
models with global Peccei-Quinn symmetry 
[\cite{PQ}].

In another 
class of models the left-handed neutrino achieves a
Majorana mass through
radiative corrections involving a scalar particle that mediates 
lepton number violating interactions [\cite{LNure}].
This scalar is not allowed to obtain vacuum expectation value, since it would
break the existing electroweak symmetry incorrectly.
The induced neutrino mass is typically 
$m_{\nu_\ell} = f^2 m_\ell$, where $m_\ell$ is the mass of the charged lepton
and $f$ is the Yukawa coupling of the charged lepton. 

The models where neutrinos are Dirac particles are in general unnatural in
the sense that the smallness of the neutrino mass
should be put in by hand, or one has to allow a large hierargy
($\geq 10^6$) between the Yukawa couplings of charged leptons and neutrinos.
In the latter case 
the right-handed neutrinos can be included as singlets as 
can be done in the $SU(2)_L\otimes U(1)_Y$ symmetric models. 
However, it is possible to construct models where we have see-saw 
like mechanism for massive Dirac neutrinos. This 
can be done e.g., in a class of models containing an $SU(n)_F$  family 
symmetry 
broken by Higgs scalars transforming according to the fundamental
representation of the gauge symmetry [\cite{DirFam}].

The models where the heavy neutrinos are mirror particles [\cite{Mirror}] are
also considered in this Thesis. Mirror fermions are particles which
couple to the ordinary $W$ boson in $V+A$ currents, in contrast with ordinary
fermions which couple in
$V-A$ currents. 
Mirror fermions arise naturally in so-called family unified models based on
large orthogonal groups [\cite{ortho}], as well as in a class of models based
on large unitary groups [\cite{unitar}] or on exceptional groups
[\cite{exept}]. Also in the Kaluza-Klein theories [\cite{kaluza}] and in some
composite models [\cite{compos}] the existence of the mirror fermions 
is predicted.

The lower bound of charged mirror fermion masses is set by the undiscovery of
these particles at LEP, and it is $m_F \gsim {\cal O}(45)$ GeV.
The lower bound on mirror neutrino masses depend heavily on the mixing of these
particles and therefore there is no generally valid limit available. 
On the other hand, they must be lighter than ${\cal O} (300)$ GeV.
This is due to the fact that their masses are generated in the spontaneous
symmetry breaking of $SU(2)_L\otimes U(1)_Y$ gauge symmetry and are therefore 
limited,
if one wants to maintain the
validity of perturbation theory by not allowing too large Yukawa couplings.
Even if mirror fermions were too heavy to be observed, they may have manifest
effects through possible mixings with the ordinary particles. As a result of
such a mixing the gauge interactions of the ordinary particles would be a
mixture of the $V-A$ and $V+A$ interactions. Our analysis in Papers I and II
takes this possibility into account by allowing for a general $V$, $A$
structure of interactions.

\vsas
\subsection{Left-Right Symmetric Model and Massive Neutrinos}   \label{3.2}

The original motivation of the left-right symmetric model 
based on an $SU(2)_R\otimes SU(2)_L\otimes U(1)_Y$ gauge symmetry
is related to the 
maximal parity violation observed in the low-energy weak interactions. In the 
SM, parity violation is arranged quite artificially by treating left- and 
right-handed fermions on unequal footing. In the left-right symmetric model 
instead, the parity violation has a dynamical origin.
The left-handed and right-handed
fields are treated on the same basis in Lagrangian,
but the vacuum state is not invariant under the parity transformation.
The parity violation is related to the spontaneous breaking of the left-right 
symmetry 
below some energy scale considerably larger than the electroweak scale.

In the left-right symmetric model the $U(1)$
generator has a clear physical meaning:  it is 
the baryon number operator $B$ minus the lepton number operator $L$
[\cite{B-L}].
As this symmetry is spontaneously 
broken, there exist lepton and baryon number violating interactions.
The lepton number violation 
manifests itself most clearly 
in neutrino physics, e.g., in the so-called neutrinoless double-$\beta$ decay
[\cite{DBDec}]. 
Lepton number violating processes could also play a crucial 
role in explaining
the baryon number asymmetry of the universe 
[\cite{MatAnm}].

The particle spectrum of the left-right symmetric model
contains all the SM particles, added with the right-handed neutrinos, and 
new heavier
gauge bosons $W_2$ and $Z_2$ associated with the $SU(2)_R$ symmetry as well as
several new Higgs bosons. The decomposition of the scalar sector depends on the
symmetry breaking chain.
One usually considers a model where
the Higgs sector consists of one bidoublet Higgs and so-called left- and 
right-handed triplet Higgses
[\cite{LRHIG}].
In some versions of the model 
the triplets are replaced by two doublet fields, or there are both doublets and
triplets.
In the most simple model, 
the Higgs sector consists of one bidoublet field and one right-handed triplet
field, enough to break the symmetry appropriately and to generate the fermion
masses. In this Thesis we will assume that 
both the left- and right-handed triplet fields are present.

Under the $SU(2)_R\otimes SU(2)_L\otimes U(1)_{B-L}$ gauge 
symmetry each fermion generation is assigned according to  
($k=\samepage 1,2,\ldots$)
\begin{equation}
\begin{array}{lccclclcccl}
L_{kL}&=&\left(\begin{array}{c} \nu_{\ell_k}\\ \ell_k\end{array}\right)_L&
\sim& (1,2,-1), &\; &
L_{kR}&=&\left(\begin{array}{c} \ell_k\\ \nu_{\ell_k}\end{array}\right)_R&
\sim&(2,1,-1),
\vspace{2.0mm}\\
Q_{kL}&=&\left(\begin{array}{c} u_k\\ d_k\end{array}\right)_L&\sim&
(1,2,\frac{1}{3}),& &
Q_{kR}&=&\left(\begin{array}{c} u_k\\ d_k\end{array}\right)_R&\sim&
(2,1,\frac{1}{3}).
\end{array}\label{LRFer}
\end{equation}
This is a natural extension to the SM particle spectrum 
(\ref{SMPSp}) treating left- and right-handed fermions equally.

The breaking of
the $SU(2)_R\otimes 
SU(2)_L\otimes U(1)_{B-L}$ gauge symmetry  
to the electromagnetic gauge symmetry $U(1)_{em}$ proceeds through two stages.
The break down chain is
\begin{equation}
SU(2)_R\otimes SU(2)_L\otimes U(1)_{B-L} \stackrel{M_R}{\longrightarrow} 
SU(2)_L\otimes U(1)_{Y} \stackrel{M_{EW}}{\longrightarrow} U(1)_{em}\, .
\end{equation}
The first spontaneous breaking at the scale $M_R$ is due to the right-handed
triplet Higgs field $\Delta_R \sim (3,1,+2)$, and
the spontaneous breaking 
at the electroweak scale $M_{EW}$ is due to the bidoublet 
$\phi\sim (2^*,2,0)$ and the left-handed triplet field $\Delta_L\sim (1,3,+2)$.
The left- and right-handed triplet and the bidoublet fields are:
\begin{equation}
\begin{array}{lclclcl}
\Delta_{L/R} &=& \left(\begin{array}{ll} \frac{1}{\sqrt{2}}\Delta^+_{L/R} & 
\Delta^{++}_{L/R}\\ \Delta^0_{L/R} & -\frac{1}{\sqrt{2}}\Delta^+_{L/R}
\end{array}\right) 
\,, &\;\;\;&
\phi         &=&  \left(\begin{array}{ll} \phi^0_1 & \phi^+_1\\
\phi^-_2&  \phi^0_2 \end{array}\right)\,,
\end{array}
\end{equation}
which at the spontaneous symmetry breaking acquire 
the following 
vacuum expectation values:
\begin{equation}
\begin{array}{lclclcl}
\langle\Delta_{L/R}\rangle &=& \left(\begin{array}{ll} 
0 &0\\ v_{L/R} & 0 \end{array}\right)
\,,&\;\;\;&
\langle\phi\rangle &=&  \left(\begin{array}{lc} \kappa_1 & 0\\  
0&  \kappa_2e^{\dot{\imath}\alpha} \end{array}\right)\, ,
\end{array}
\end{equation}
where $v_{L/R}$, $\kappa_1$, $\kappa_2$ and $\alpha$ are real parameters.

At the first stage  of the symmetry breaking  
the right-handed vector bosons $W_R$ and $Z^0_R$ acquire the masses 
\begin{equation}
\begin{array}{lclclcl}
M_{W_R} &=& gv_R\,, 
&\;\;\;\;& M_{Z_R} &=& M_{W_R} 
\sqrt{\frac{\displaystyle 2\cos^2\theta_W}{\displaystyle
\cos 2\theta_W}}\,,\label{ZRBos}
\end{array}
\end{equation}
where $\theta_W$ is the weak mixing angle of the left-right symmetric 
model. 
The relation (\ref{ZRBos}) is valid only if we assume $g_L = g_R$,
i.e. that the gauge coupling constants of the left- and right-handed
sectors are equal. This is a common assumption,
but in most of the results presented in this Thesis, it is not made.

The vacuum expectation values of the left-handed triplet and the bidoublet
fields
force the symmetry breaking at the electroweak scale.
The experimental results [\cite{rhopar}] on
the parameter
\begin{equation}
\rho_0 = \left(\frac{M_W}{\cos\Theta_W M_Z}\right)^2
= \frac{\sum_{\rm isospin}\left\{{\left[T^i(T^i+1)-(T^i_3)^2\right]\langle H_i
\rangle}\right\}}{\sum_{\rm isospin}\left\{{2(T^i_3)^2\langle H_i\rangle}
\right\}}
\, ,
\end{equation}
where $T$ and $T_3$ are the $SU(2)_L$ isospin and its third component,
respectively, and the sum goes over the Higgs fields $H_i$ that acquire a
non-vanishing vacuum expectation value,
imply that the vacuum expectation values 
$\kappa_1$, $\kappa_2$ and 
$v_L$ must obey $v_L \ll \kappa_1,\kappa_2$.
Therefore, the main contribution to the mass of the left-handed boson comes
from the vacuum expectation value of the bidoublet field.

Since the bidoublet field $\phi$
transforms nontrivially under both the $SU(2)_L$ and $SU(2)_R$
groups, the left- and right-handed gauge bosons may mix with each other. 
The physical eigenstates of the vector bosons are hence in general given by
\begin{equation}
\begin{array}{lcl}
W_1 &=& \cos\zeta W_L + e^{\dot{\imath}\alpha}\sin\zeta W_R\simeq W_L, \\
W_2  &=& -e^{\dot{\imath}\alpha}\sin\zeta W_L + \cos\zeta W_R\simeq W_R,\\
A   &=& \sin\theta_W\left(W^3_L+W^3_R\right)+\sqrt{\cos 2\theta_W}\, B
\equiv \gamma,\\
Z_L &=& \cos\theta_WW^3_L-\sin\theta_W\tan\theta_WW^3_R-\tan\theta_W
\sqrt{\cos 2\theta_W}\, B\simeq Z_1,\\
Z_R &=& \sec\theta_W\sqrt{\cos 2\theta_W}\,W^3_R-\tan\theta_W B\simeq Z_2
\end{array}
\end{equation}
where $A$ is the massless gauge field of 
the unbroken $U(1)_{em}$ symmetry, the photon, and
\begin{equation}
\tan\zeta = \frac{\kappa_1\kappa_2}{\kappa_1^2+\kappa_2^2 + 8v^2_R}\, .
\end{equation}
The masses of the light weak bosons are given approximately by
\begin{equation}
\begin{array}{lcl}
M^2_{W_L}&\simeq & \frac{\displaystyle 1}{\displaystyle 2}g^2
\left(\kappa^2_1+\kappa^2_2 \right) \equiv M^2_{W_1}\, ,\\
M^2_{Z_L} &\simeq& M^2_{Z}\left[1-\eta A_W\right] + {\cal O}(\eta^2)\, ,
\end{array}\label{LRBoMa}
\end{equation}
where $A_W \equiv \frac{1}{2}\cos 2\theta_W({1-\frac{1}{4}
\tan^4\theta_W})$ and $\eta = M^2_{W_L}/M^2_{W_R}$.
By comparing equations (\ref{LRBoMa}) and (\ref{SMBosMa}) one can see that 
$M_{Z_L}$
is always less than the mass $M_Z$ of the $Z$-boson of the SM.

The charged and neutral current Lagrangians are given by
\begin{equation}
\begin{array}{lcr}
{\cal L}^{CC}_{wk}&\!\simeq\!&\frac{\displaystyle g}{\displaystyle 2\sqrt{2}}
\left[{\left({\cos\zeta\,{J^+_L}_\mu+ e^{\dot{\imath}\alpha}\sin\zeta\,
{J^+_R}_\mu}\right){W^+_L}^\mu}\right.\;\;\;\;\;\;\;\;\;
\;\;\;\;\;\;\;\;\;\;\;\;\;\;\;\;\;\;\;\;\;\;\;\;\;\;\;\;\;\;\;\;\;\;\;\;\;\;\\
&&+\left.{ 
\left({\cos\zeta\,{J^+_R}_\mu} - e^{\dot{\imath}\alpha}\sin\zeta\,
{J^+_L}_\mu\right){W^+_R}^\mu+ {\rm h.c.}}\right]
\;\;\;\;\;\;\;\;\;\;\;\;\;\;\;\;\;\;\;\;\;\;\;\;\;\;\;\;\;\;\;\;\\
{\cal L}^{NC}_{wk}&\!=\!&eJ^{em}_\mu A^\mu+
\frac{\displaystyle g}{\displaystyle\cos\theta_W}
\left\{{\left[{{J^Z_L}_\mu\!-\eta\cos\theta_W\!\left({\sin^2\theta_W{J^Z_L}_\mu
\!+\cos^2\theta_W{J^Z_R}_\mu}\right)}\right]\!Z_L^\mu}\right.\;\;\;\;\\
& &\!+\left.{\left({\cos^2\theta_W}\right)^{-1/2}\left({\sin^2\theta_W
{J^Z_L}_\mu\!+\cos^2\theta_W{J^Z_R}_\mu }\right)\!Z_R^\mu }\right\}\, ,
\end{array}\label{HeffCC}
\end{equation}
where the weak neutral currents are $J_{L/R}^Z=J_{L/R}^3-Q\sin^2
\theta_W\, J^{em}$ and the $J^{em}$ is the electromagnetic current.
The observed $V-A$ form of the weak interactions at low energies
follows from the condition $\eta\ll 1$ and $\zeta\ll 1$ [\cite{zetaeta}].

The most general Yukawa coupling Lagrangian of leptons is
\begin{equation}
{\cal L}_Y = \bar{L}_{iL} ({h_{ij}\phi+
\tilde{h}_{ij}\tilde{\phi}}) L_{jR} +
(f_L)_{ij}\bar{L}^c_{iR}(\dot{\imath}\tau_2)\vec{\tau}
L_{jL}\cdot{\stackrel{\rightarrow}{\Delta}}_L
 +(f_R)_{ij}\bar{L}^c_{iL}(\dot{\imath}\tau_2)\vec{\tau}
L_{jR}\cdot{\stackrel{\rightarrow}{\Delta}}_R
+ {\rm h.c.}\label{LRYbd}
\end{equation}
where $h$, $\tilde{h}$, and $f_L(R)$ 
are the $n\times n$ Yukawa coupling matrices ($n$ is the number of lepton 
generations) and $\tilde{\phi} = \tau_2 \phi^* \tau_2$ is the conjugated 
bidoublet field. The most general Yukawa coupling Lagrangian of quarks is 
similar to (\ref{LRYbd}), but there are no couplings to the triplet fields
because of the $B-L$ symmetry.

\vsas
\subsubsection{See-saw Mechanism}   \label{3.2.1} 

Let us consider more closely the generation of neutrino masses in the $SU(2)_R
\otimes SU(2)_L\otimes U(1)_{B-L}$ model.
According to eq. (\ref{LRYbd}), neutrino fields acquire in the symmetry
breaking Dirac masses due to 
their coupling with the bidoublet Higgs $\phi$,
\begin{equation}
{\cal L}^D_{m_\nu} = \left(h_{ij}\langle\phi^0_1\rangle
+ \tilde{h}_{ij}\langle{\phi^0_2}^*\rangle\right)\bar{\nu}_{iL}\nu_{jR}+
{\rm h.c.}\, ,
\end{equation}
and Majorana masses due to their coupling with the triplet Higgses $\Delta_L$
and $\Delta_R$,
\begin{equation}
{\cal L}^M_{m_\nu}=(f_L)_{ij}\langle\Delta^0_L\rangle\bar{\nu}_{iR}^c
\nu_{jL}
+ (f_R)_{ij}\langle\Delta^0_R\rangle\bar{\nu}_{iL}^c\nu_{jR} + {\rm h.c.}\, .
\end{equation}
The complete mass Lagrangian can thus be written in the form
\begin{equation}
{\cal L}_{m_\nu} = \left(\bar{\nu}_{iR}^c\,\bar{\nu}_{iR}\right)\!\! \left(\!
\begin{array}{cc} (f^*_L)_{ij}\langle{\Delta^0_L}^*\rangle & 
\frac{1}{2}\left(
h_{ij}^*\langle{\phi^0_1}^*\rangle+ {\tilde{h}}_{ij}
\langle{\phi^0_2}\rangle\right)\\
\frac{1}{2}\left(h_{ji}^*\langle{\phi^0_1}^*\rangle+ \tilde{h}_{ji}\langle
{\phi^0_2}\rangle\right) & (f_R)_{ij}\langle\Delta^0_R\rangle\\
\end{array}\!\right)\!\!\left(\begin{array}{c} 
\!\nu_{jL}\! \\ \!\nu^c_{jL}\!\end{array}\right) + {\rm h.c.},\label{NeuMAL}
\end{equation}
where we have used the relation $\bar{\nu}_R\nu_L = \frac{1}{2}\left(
\bar{\nu}_R\nu_L+\bar{\nu}^c_R\nu^c_L\right)$.
In order to find the physical neutrino fields one has to  
diagonalize this Lagrangian.

As already mentioned, the
experimental constraints require that $\langle\Delta^0_R\rangle \gg 
\langle\phi^0_{1,\,2}\rangle \gg \langle\Delta^0_L\rangle$. 
It is therefore reasonable to approximate $(f^*_L)_{ij}\langle{\Delta^0_L}^*
\rangle \simeq 0$ in (\ref{NeuMAL}).
The mass Lagrangian
then obtains the following see-saw form:
\begin{equation}
{\cal L}_{m_\nu} 
\approx
\left(\bar{\nu}_R^c\,\bar{\nu}_R\right)\! \left({
\begin{array}{cc}\!\!0 &\!m_D\\
\!m^T_D &\!\!m_R
\end{array}}\right)\!\!\left(\begin{array}{c} 
\!{\nu_L}\! \\ \!{\nu^c_L}\!\end{array}\right) \! + \! {\rm h.c.}
\equiv \left(\bar{\nu}_R^c\,\bar{\nu}_R\right) M
\left(\begin{array}{c}  \!{\nu_L}\! \\  \!{\nu^c_L}\!\end{array}\right)
 \! + \! {\rm h.c.}\, , \label{LagSS}
\end{equation}
where 
the expressions of the $3\times 3$ matrices $m_D$ and $m_R$ can be read from
(\ref{NeuMAL}).
The left-handed neutrino fields appearing here
can be identified with 
the left-handed neutrinos $\nu_{eL}$, $\nu_{\mu L}$ and $\nu_{\tau L}$.
In the case of one neutrino flavor the mass Lagrangian can be 
diagonalized with an unitary matrix transformation $U$ in the following way:
\begin{equation}
U^\dagger MU = \left(\!\!{\begin{array}{cc}\cos\theta &\sin\theta\\-\sin\theta&
\cos\theta\end{array}}\!\!\right) \!\! \left(\!\!{\begin{array}{cc}0 &m_D\\
m_D &m_R\end{array}}\!\!\right)\!\!\left(\!\!{\begin{array}{cc} \cos\theta
&-\sin\theta\\ \sin\theta&\cos\theta\end{array}}\!\!\right) = \left(\!\!
{\begin{array}{cc}\widehat{m}_1 & 0\\
0&\widehat{m}_2\end{array}}\!\!\right) = M'\, .
\end{equation}
The  eigenvalues of the matrix $M'$ ($m_R\gg m_D$) are:
\begin{equation}
\widehat{m}_{1,2} = \frac{1}{2}m_R\left[1\pm\sqrt{1+\left({\frac{m_D}{m_R}}
\right)^2}\right] \simeq \left\{\begin{array}{l}\;\;m_R \\ 
\!\!-\frac{\displaystyle m^2_D}{\displaystyle m_R}
\end{array}\right.\,.\label{MaMAne}
\end{equation}
They are related to the masses of the physical neutrinos via $m_i=\widehat{m}_i
\eta_i^{CP}$, where $\eta_i^{CP}=\pm 1$ is the CP parity of the neutrino
fields. The corresponding mass eigenstates $\chi_1$ and $\chi_2$ are given by
\begin{equation}
\begin{array}{lcl}
\chi_1 &\simeq& \nu_L + \nu^c_R - \frac{\displaystyle m_D}{\displaystyle m_R}
\left(\nu_R+\nu^c_L\right),
\\
\chi_2 &\simeq& \frac{\displaystyle m_D}{\displaystyle m_R}\left(\nu_R+
\nu^c_L\right) - \nu_R + \nu^c_L\, . 
\end{array}
\end{equation}
As seen from this, the physical neutrinos $\chi_i$ are
Majorana particles, i.e.
they fulfill the Majorana condition $\chi_i = \chi^c_i,\; i=1,2$.
The neutrino mixing angle $\theta$ is given by
$
\tan2\theta \simeq {2m_D}/{m_R}.
$
The current sates of the neutrino fields in terms of the mass eigenstates are
\begin{equation}
\begin{array}{lcl}
\nu_L &\simeq & -\chi_{2L} + \frac{\displaystyle m_D}
{\displaystyle m_R}\chi_{1L},
\\
\nu^c_L &\simeq & \chi_{1L} + 
\frac{\displaystyle m_D}{\displaystyle m_R}\chi_{2L}\, .
\end{array}
\end{equation}
That is, the left-handed neutrino $\nu_L$ 
appearing in low-energy phenomena corresponds to the light
mass eigenstates $\chi_2$ and the right-handed 
neutrino $\nu_R$ ($\nu^c_L$) 
corresponds to the heavy mass eigenstate $\chi_1$.

\subsection{Heavy Neutrino Physics
at Future $e^+e^-$ Colliders}\label{3.3}

Future high-energy $e^+e^-$ colliders will offer a clean environment for
investigating the properties of heavy neutrinos. This is an important issue
since heavy neutrinos are 
one of the main characteristics of many extended gauge models that give the SM
as a low-energy approximation.

In Papers I and II we investigate the production of heavy neutrinos in
$e^+e^-$ collisions. In Paper I the main focus is on the collision energies
$\sqrt{s} = 150 - 300$ GeV, i.e. in the energy range of the LEP200 and a low
energy linear collider. The
production cross sections of the pair production of heavy neutrinos
($e^+e^-\rightarrow  \overline{N}_1N_2$) and a single heavy neutrino production
($e^+e^-\rightarrow \overline{N}_1\nu_2$ or $N_1\overline{\nu}_2$), as well as
the decay properties of heavy neutrinos, are analyzed.
As a model for weak interactions we use phenomenological model
which has one set of weak vector bosons with general (real) $V\pm A$
couplings. The choice of just one vector boson generation is justified by
the experimental results 
[\cite{zetaeta,WRLIMIT}], 
that the mass of the heavier
vector bosons is much higher than the chosen collision energy scale.
In our analysis both Majorana and Dirac cases are studied and
especially the differences between heavy Dirac and Majorana neutrino
production, as well as the differences between the $V+A$ and $V-A$ currents,
are investigated. 

The results we obtained in Paper I show that the pair production rate of heavy
Majorana neutrinos is smaller than the corresponding rate of Dirac neurinos
with the same mass. The total pair production rate is the same for both
left- and right-handed neutrinos, but the angular distributions are different
and one should be able to use forward-backward asymmetries for distinguishing
right- and left-handed neutrinos provided one has an opportunity to collect
enough events. The angular distributions can be used also to distinguish
the Dirac and Majorana neutrinos.

In the single heavy neutrino production the Majorana neutrino production rate
is two times bigger that the Dirac neutrino production rate. Provided the
coupling constants are equal for both $V+A$ and $V-A$ currents, the 
production rates are typically by a 
factor of $1.5$ larger in the case of $V+A$ currents.
Since the angular distributions are dramatically different in the single heavy
neutrino production the separation of Dirac and Majorana neutrinos, as well as
the right- and left-handed neutrinos, is possible with considerably lower
statistics than in the pair production.
The experimental signal for single heavy neutrino production consists of
heavily unbalanced events with two charged leptons or unbalanced hadron jets
accompanied with one charged lepton. These very spectacular events
should make the detection of heavy neutrinos easy because the
expected background is small.

In Paper I the subsequent decays of the heavy neutrinos are also studied.
The results verify the well known fact
that the total decay width of a Majorana neutrino is two times bigger that
of the Dirac neutrino with the same mass and the same couplings.
The angular distributions of leptons
produced in the decay are also different for Dirac and Majorana neutrinos.
Based on these two properties we propose a 
powerful and experimentally sound method to distinguish the Majorana and Dirac
cases via studying the correlated production of the same sign lepton pairs from
the $\overline{N}N$ system. Also the differences in angular distributions and
lifetimes can be used for distinguishing heavy Dirac and Majorana
neutrinos, but it would require higher statistics.

In Paper II the production of heavy Dirac and Majorana neutrinos in $e^+e^-$
collisions ($e^+e^-\rightarrow \overline{N}_1N_2$) are studied using a gauge
model with an arbitrary number of neutral vector bosons $Z^0_i$, $i=1,\ldots, 
N$ and charged vector bosons $W^\pm_j$ $j=1,\ldots, M$ with the most general
$V$, $A$ couplings. The production of heavy neutrinos in Higgs exchange
channels are also analyzed since in specific models some of the Yukawa
couplings could be large enough in these channels to give measurable
contribution to the production rate.
Numerical results are presented for the case of a minimal left-right
symmetric model with two vector boson generations and with a minimal set of
Higgs fields, i.e. with two triplets and one bidoublet. The
model is simplified by using a phenomenologically motivated set of gauge and
Yukawa couplings.
A similar analysis with a more strictly defined version of the left-right
symmetric model is presented in [\cite{hehio}].

The main issues of Paper II are to study carefully the interference patterns
between different production channels, as well as to find differences between
left- and right-currents and between Dirac and Majorana neutrinos.
By separating the interference terms we found that they can cause big
changes to the heavy neutrino production threshold behaviour, as well as to
the production rates in the vicinity of the heavy boson pole. The interference
terms are non-negligible although some of the interfering channels are much
smaller that the other channels.
In the model specified in Paper II some $50$ to $100$ heavy Dirac or Majorana
neutrinos with a mass $m_N=150$ GeV are expected to be produced annually at a
$500$ GeV linear collider assuming a luminosity ${\cal L}_{\rm year} = 10$
${\rm fb^{-1}}$. 
It is also shown that, in order to make a reliable distinction between Dirac
and Majorana cases, higher statistics would be required.
The possibility to do production threshold scans would also make the
distinction of the Dirac and Majorana neutrinos easier, but for this
one would need a linear collider with adjustable electron and positron beam
energies and a very flexible focusing systems.

Concerning the angular distributions and even the decays of the heavy 
neutrinos, the arguments presented for Paper I are also valid for Paper II.
There
are also differences due to the more general model used in Paper II, e.g, in
the pair production of heavy neutrinos, the pole of the second neutral boson
$Z_2$ enhances the pair production rate drastically. This is due to the fact
that heavy neutrinos tend to have non-suppressed coupling to the heavier
bosons. This is not true in the single heavy neutrino production since either
the heavy neutrino or the light neutrino couplings to the charged vector bosons
are suppressed. One should also note that among all heavy neutrino pair
production channels the Higgs boson exchange processes give always the smallest
contribution. On the other hand, the results in Paper II also indicate that the
production of heavy neutrinos produced via a single charged triplet Higgs
exchange would be visible only at high collision energies
($\sqrt{s} = 1 - 2$ TeV) or at
$e^+e^-$ linear collider
with a very high luminosity (${\cal L}_{\rm year} \ge 100$ ${\rm fb^{-1}}$). 

\clearpage

\vsas
\section{Conclusions}   \label{4}

Although high-energy physics experiments have increased our knowledge
about particle physics interactions and phenomena to a remarkable level,
it is by no means excluded that new phenomena are found when one goes beyond
the energy and precision reach of our present facilities.
Especially the future collider experiments, which aim to the TeV energy scale,
have a potential to reveal phenomena unknown at present.
It will be of great importance to carefully study in advance the new physics
topics one could explore with these next generation facilities.
The aim of this Thesis has been to study the neutrino physics phenomenology 
at the future $e^+e^-$ colliders (research Papers I, II and III) and 
to investigate the phenomenology of the most poorly known Standard Model
constituents, the top quark and the tau neutrino, at $e^+e^-$ colliders
(Papers III and IV). 

The goal of Paper III is to find a method to observe the only directly
undiscovered fermion of the SM, the tau neutrino. 
The proposed method
consists of two parts. The first part is a very asymmetric $e^+e^-$ collider
which runs either at the $Z^0$-pole, i.e. at $\sqrt{s}=M_{Z^0}$, or at 
$\sqrt{s}\approx4.2$ GeV, i.e. just above the $\tau$-lepton pair production
threshold. The second part consist of a simple rejection detector 
(veto-trigger) around the interaction point and of a coarsely instrumented
muon spectrometer with a large volume and mass. The asymmetric collider could
be built by using a positron beam of the TeV range linear collider as a high
energy beam, and the characteristics of the low energy electron beam could be 
similar to those of electron damping rings needed for the future $e^+e^-$
linear colliders. Using the realistic accelerator and detector setup proposed
in Paper III, the discovery signal is found to be quite marginal.
While making the discovery of the tau neutrino possible the
apparatus would not provide a possibility to a very systematic study of the
properties of the produced tau neutrinos.
On the other hand, with these minimal specifications one should be able to
build the whole apparatus, alongside an existing linear collider, with a
fraction of the cost of a stand alone experiment with specialized accelerator
and detectors.

The single top production considered in Paper IV will be an important process
in the next generation $e^+e^-$ linear colliders. 
According to present knowledge, at 
LEP200 the cross section will be, however, too small to yield an observable
signal, in contrast to the more optimistic estimates presented in Paper IV.

One interesting prediction common to many extensions of the SM is the existence
of massive neutrinos. The main attention in this Thesis is paid on the
left-right symmetric model which produces via the see-saw mechanism in each
generation a light and a heavy neutrino which are Majorana particles. The
observation of a heavy neutrino in $e^+e^-$ collisions would prove that the SM
is not the final theory, since it does not have such particles.
Due to their small mass, it is very difficult to probe the nature of light
neutrinos in scattering experiments. In the processes involving heavy
neutrinos, instead, the signals which would make distinction between Dirac and
Majorana neutrinos would be easier to detect. The results presented in Paper
I and, especially, in Paper II should facilitate such investigations by giving
the relevant cross section and decay widths in a form which applies to a large
class of extended models.

\clearpage

\setlength{\baselineskip}{.6cm}

\vsas

\end{document}